\documentclass[apj]{emulateapj}

\usepackage{multirow}
\usepackage{ulem}
\usepackage{color}
\usepackage{url}
\def\rhessi{{\textit{RHESSI}}}

\def\mw{{microwave}}

\def\gs{{gyrosynchrotron}}

\def\peak{{ _{\rm peak}}}

\received{\today}
\revised{\today}
\accepted{\today}

\shorttitle{3D flare modeling}
\shortauthors{Fleishman et al.}

\begin{document}


\title{Revealing evolution of nonthermal electrons in solar flares using 3D modeling }

\author{Gregory D. Fleishman$^1$}

\author{Gelu M. Nita$^1$}

\author{Natsuha Kuroda$^1$}

\author{Sabina Jia$^2$}

\author{Kevin Tong$^2$}

\author{Richard R. Wen$^2$}

\author{Zhou Zhizhuo$^2$}

\altaffiliation{$^1$Physics Department, Center for Solar-Terrestrial Research, New Jersey Institute of Technology
Newark, NJ, 07102-1982}

\altaffiliation{$^2$Union County Vocational-Technical Schools, 1776 Raritan Road, Scotch Plains, NJ, USA, 07076-2997}

\begin{abstract}
Understanding nonthermal particle generation, transport, and escape in solar flares requires detailed quantification of the particle evolution in the realistic 3D domain where the flare takes place. Rather surprisingly, apart of standard flare scenario and integral characteristics of the nonthermal electrons, not much is known about actual evolution of nonthermal electrons in the 3D spatial domain. This paper attempts to begin to remedy this situation by creating sets of evolving 3D models, the synthesized emission from which matches the evolving observed emission. Here we investigate two contrasting flares: a dense, ``coronal-thick-target'' flare SOL2002-04-12T17:42, that contained a single flare loop observed in both \mw\ and X-ray, and a more complex flare, SOL2015-06-22T17:50, that contained at least four distinct flaring loops needed to consistently reproduce the \mw\ and X-ray emission. Our analysis reveals differing evolution pattern of the nonthermal electrons in the dense and tenuous loops; however, both of which imply the central role of resonant wave-particle interaction with turbulence. These results offer new constraints for theory and models of the particle acceleration and transport in solar flares.
\end{abstract}


\section{Introduction}


Creating a complete picture of particle acceleration in solar flares requires detailed knowledge of where the particles are accelerated and how they evolve in the actual three-dimensional (3D) domain, where the flare happens. However, such information is not routinely available since it cannot easily be derived from the available data sets nor modeled from the first principles. Indeed, most information about the nonthermal electrons is derived from X-ray emission which is produced due to collisions of nonthermal electrons with target particles (mainly, protons) via bremsstrahlung. For example,  \textit{Reuven Ramaty High Energy Solar Spectroscopic Imager} \citep[\rhessi,][]{lin2002} produces both spectral and imaging information and, thus, is the most appropriate instrument with which to study electron acceleration and transport. However, the limited dynamic range of \rhessi\ makes it hard or impossible to detect hard X-ray (HXR) emission from the coronal part of the flaring loops against much brighter footpoints. An exception is dense, the so-called ``coronal-thick-target'' flares, where HXR emission comes from the corresponding coronal flaring loop \citep{Xu_etal_2008, 2012A&A...543A..53G, 2012ApJ...755...32G, 2013ApJ...766...28G,
2013ApJ...769L..11L}. But even in such cases, the \rhessi\ imaging is typically made over rather long time (e.g., half a minute or longer), so the detailed dynamics cannot be captured. In addition, study of the electron acceleration in the coronal-thick-target flares is biased by the cases of rather dense sources, while cannot help to reveal the complete picture of the electron acceleration and transport in the whole variety of the tenuous and dense flaring flux tubes, although both tenuous and dense acceleration regions were reported \citep{Fl_etal_2011, Fl_etal_2016_narrow}. A viable enhancement of the nonthermal electron probing comes from the microwave imaging spectroscopy \citep{Gary_etal_2013}. But even the immediate outcome of the \mw\ data analysis will be the line-of-sight (LOS) integrated distributions of the parameters in the image plane, rather than the true 3D distribution. Therefore, in addition to the observational data set, revealing the true 3D structure must rely on the realistic data-constrained modeling. 

In this paper we attempt to constrain the flare dynamics in the 3D domain by creating evolving 3D models capable of producing synthetic \mw\ emission that matches the observed one during a time range. To do so we start from ``master'' models, developed using GX Simulator \citep{Nita_etal_2015} earlier for single time frames and validated by comparison with all available \mw\ and X-ray constraints \citep{Fl_Xu_etal_2016, 2018ApJ...852...32K}. Then, assuming that the basic flare topology does not change drastically over some time range, we use these master models and alter  some of the physical parameters of them to derive model set, each of which is capable of producing the synthesised \mw\ spectrum that matches the observed one at a given time frame.   This set of the ``snapshot'' models is then adopted to represent the evolving 3D model of the flare. It is apparent that the evolving flare model obtained this way is not unique, while a family of models could be consistent with the applied observational constraints. In particular, the magnetic structure of the evolving model is fixed by construction, which is of course an approximation to the reality. However, given that the magnetic loop structure is fixed, the evolving distribution of the nonthermal electrons offers one of the likely evolutionary pattern of the physical parameter variation, main trends of which truthfully reproduce the actual parameter trends.    

We note that to address the evolution we attempt to match the \mw\ spectra only, rather than both spectral and imaging data, because the spectra are known with a higher cadence than the images. In addition, we do not see any substantial fast evolution in the images, which would justify changing the magnetic topology in the model (e.g., altering the central, ``reference'' field line defining the flaring flux tube). However, after advancing our model over a certain time range (typically, 2--3 minutes) it was not possible to obtain a good match with the \mw\ data within the fixed topology; we interpret this as an indication of the magnetic topology change. Once the evolving models have been obtained, it is instructive to straightforwardly analyze trends in evolving parameters and link these trends with possible acceleration models and transport regimes, which is we discuss in some detail in Section~\ref{S_Discuss}.

We consider two flares with rather contrasting properties. The first of them, a dense, ``coronal-thick-target'' flare SOL2002-04-12T17:42, showed a rather simple, single-source morphology in both X-ray and \mw\ images.  This flare has been successfully modeled using a single asymmetric flare loop \citep[][in fact, two slightly different flaring loops were needed to model two distinct time frames--one at the rise phase and the other one at the peak phase]{Fl_Xu_etal_2016}.
The other one, SOL2015-06-22T17:50, appeared much more complex and contained at least four distinct flaring loops needed to consistently reproduce the \mw\ and X-ray emission. Only two of those four loops contributed noticeably to the \mw\ emission with the main contribution coming from a large, tenuous, ``overarching'' loop that served as an efficient trap for the nonthermal electrons accelerated presumably somewhere lower in the corona as follows from the X-ray morphology.

In what follows we discuss the trends of the physical parameters in both cases, make comparison between the trends, and formulate new constraints on the electron acceleration models and transport regimes.

\section{Single Loop Flare}
\label{S_dense_flare}

A list of coronal thick-target flares was identified and studied by \citet{Xu_etal_2008} although a few examples of such flares were reported by \citet{2004ApJ...603L.117V}. These flares appear as single loop-shaped sources at the HXR and sometimes also observed in \mw\ \citep{2013ApJ...769L..11L, Fl_Xu_etal_2016}. In particular, \citet{Fl_Xu_etal_2016} studied one of the flares from the \citet{Xu_etal_2008} list, SOL2002-04-12T17:42 with X-ray and \mw\ data augmented by 3D modeling using GX Simulator \citep{Nita_etal_2015} and found that both radio and X-ray data are well reproduced within a 3D model involving only one asymmetric loop, although the loop itself evolved from the rise to peak phase of the flare.

\subsection{Overview of SOL2002-04-12T17:42}
\label{S_overview_20020412}

SOL2002-04-12T17:42 was observed by \rhessi\ in X-rays and OVSA in \mw s. \citet{Fl_Xu_etal_2016} analyzed in detail and produced 3D models\footnote{The models and associated data files are freely available from the 3D GX model repository at $\url{http://www.ioffe.ru/LEA/SF\_AR/models/3dmodels.html}$.}
for two time frames of this flare, one at the rise phase, $\sim$17:41:58~UT, and the other at the peak phase, $\sim$17:45:10~UT. In both cases, a linear force-free field extrapolation has been performed to create a flaring flux tube, to populate it with a thermal plasma and nonthermal electrons, and validate the model via comparison of the synthetic X-ray and \mw\ images and spectra. It turned out that the magnetic structure evolved noticeably between the two time frames, so two distinct magnetic flux tubes were needed to model those two time frames.

Another interesting finding was that the \mw\ images were displaced compared with the X-ray ones for both time frames. However, 3D modeling proved that this displacement is due only to the flaring loop asymmetry such as one leg of the loop was brighter in the X-rays, while the other one--in the \mw s. It is also interesting that the flaring flux tube was noticeably thicker (the loop cross-sectional radius was bigger) at the peak time compared with the rise phase.

\citet{Fl_Xu_etal_2016} modeled the main spectral peak of the \mw\ emission with the mentioned flaring flux tubes, while the \mw\ spectrum displayed a secondary spectral peak at low frequencies, $f=1.2-2$~GHz, which was ignored at that time. It is now established that such a secondary low-frequency spectral peak, if not a coherent / plasma emission, can either be associated with a distinct large-scale ``plume'' \citep{Fl_etal_2016, Fl_etal_2017} or produced by resonant transition radiation (RTR) due to interaction of nonthermal electrons with plasma density inhomogeneities \citep{RTR, Nita_etal_2005}.

\begin{figure}[!b]
    \centering
    \includegraphics[width=0.98\columnwidth,clip]{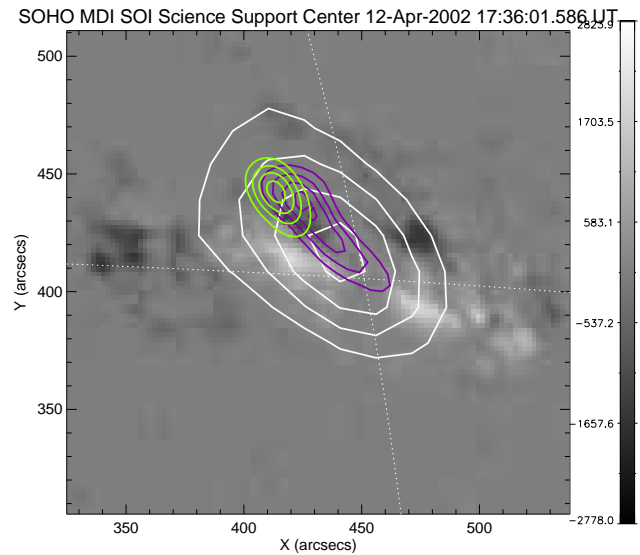}\\
    \caption{\label{f_20020412_images} HXR and \mw\ images on top of MDI LOS magnetogram obtained at $\sim$17:36~UT: 12-25~keV image obtained over one minute starting 17:45:48~UT (green), and frequency-time \mw\ images obtained over 64~sec starting 17:45:46~UT at 2.8-5.6~GHz (violet) and 1.2-1.8~GHz (white). Their projected locations are consistent with the emission coming from the same magnetic loop.  }
\end{figure}

To make a choice between those possibilities, here we produced the flare images at low \mw\ frequencies, $f=1.2-2$~GHz, and plotted them along with other \mw\ and HXR images on top of \textit{SoHO}/MDI LOS magnetogram, see Figure~\ref{f_20020412_images}. It is apparent from the figure that the low-frequency source projects onto the same solar area as the HXR and higher-frequency \mw\ images. This spatial relationship and modest size of the low-frequency source favor the RTR nature of this low-frequency component \citep[cf.][]{Nita_etal_2005}, rather than the \gs\ origin from a large loop as in \citet{Fl_etal_2016, Fl_etal_2017}. We will return to implication of this finding later.

\subsection{Rise Phase of SOL2002-04-12T17:42}
\label{S_rise_20020412}

We started our modeling from the master model developed by \citet{Fl_Xu_etal_2016} for a time frame $\sim$17:41:58~UT at the rise phase of this flare. Given the central role of the \mw\ spectra for our modeling, we inspected the quality of the corresponding OVSA data\footnote{Available in the form of .ref-files directly importable by GX Simulator from the same 3D GX model repository at $\url{http://www.ioffe.ru/LEA/SF\_AR/models/3dmodels.html}$ 
or, in various formats, from the official OVSA web-site directly.} a few minutes before and after this master time frame. We found that three highest frequency channels show large fluctuations in time, while the low-frequency spectral range, $f=1.2-2$~GHz, displays an additional spectral component. Therefore, to objectively use the model-to-data residual, computed within the GX Simulator tool, we removed both the low- and high- frequency OVSA channels to create the new reference file containing 31 frequencies spanning the $2.4-14$~GHz range; for consistency we have done so for both rise and peak (Section~\ref{S_peak_20020412}) phases of this event. After removing those outliers, it became possible to further fine tune the model parameters to get a closer model-to-data match than in the \citet{Fl_Xu_etal_2016} paper, so we use this slightly modified model 
as the master one, and advanced it backward and forward in time. 

An important question is what physical parameters are to be adjusted to ensure the desired, in accordance with observations, modification of the \mw\ spectrum. Here, we consider the main factors affecting the optically thin and thick parts of the \mw\ spectrum. For example, changes of the optically thick intensity requires  \citep[see Eq. 1 in][]{Fl_etal_2017} either a change of the source area (due to nonthermal electron redistribution within a given flux tube or change of the cross-sectional radius of the flux tube)  or of the effective energy of the emitting nonthermal electrons (due to change of the low-energy cutoff, spectral index, or the magnetic field, which, for a preselected magnetic flux tube, can be emulated by a redistribution of the nonthermal electrons in the loop), while the optically thin part is primarily controlled by the nonthermal electron energy spectrum (the spectral index and high-energy cut-off). While fine tuning the parameters, we attempted to minimize the number of free parameters to change and look for coherent patterns in the parameter behaviour.

At the rise phase we managed to successfully model four minutes time interval (two minutes before and two minutes after the ``master'' time frame) with the 4~s cadence, during which two groups of physical parameters were adjusted--the 3D spatial distribution of the nonthermal electrons and the energy distribution of the nonthermal electrons, which we describe sequentially here, even though both groups affect the resulting spectrum simultaneously.

The adjustments in the 3D spatial distribution were needed primarily to change the source area required to account for the change (increase) of the \mw\ flux in the optically thick part of the spectrum. Whenever possible, we achieved the required change by a redistribution of the nonthermal electrons along the existing modeling loop with a fixed cross-sectional reference radius.

\begin{figure}
    \centering
    \includegraphics[width=0.98\columnwidth,clip]{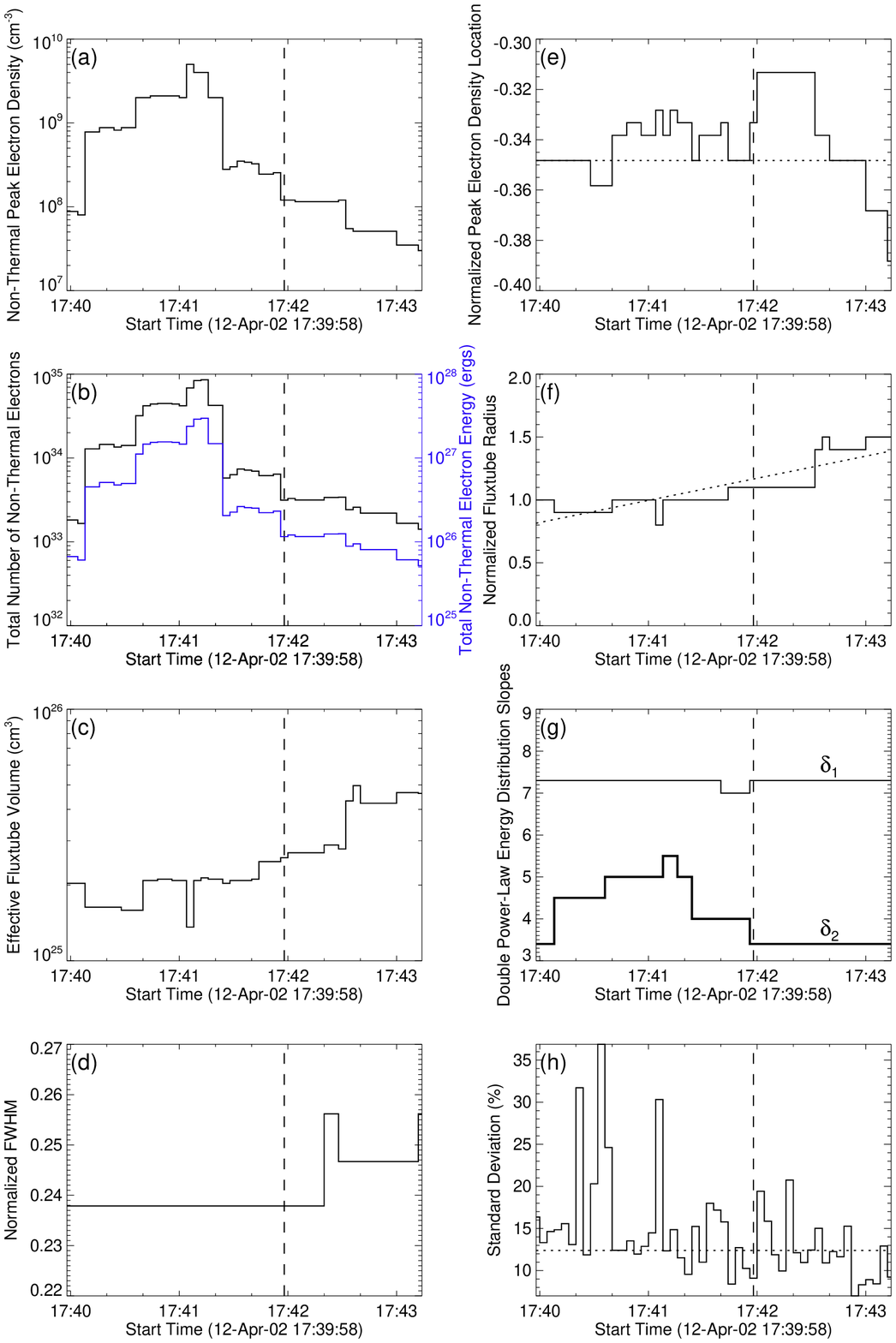}\\
    \caption{\label{parms20020412rise} Rise Phase of the SOL2002-04-12T17:42 flare: Evolution of the modeled parameters describing the spatial and energy nonthermal electron distributions around the local peak time 17:41:58 UT, which is marked by vertical dashed lines in each panel. a) Evolution of the Gaussian distribution full maximum ($n_{nth}$ in Equation~\ref{nb}); b) total number $N_{nth}$ and total energy of the nonthermal electrons; c) effective volume of the fluxtube; d) Normalized FWHM ($2\sqrt{\ln2}/q_0$) evolution,  controlled by parameter $q_0$ in Equation~\ref{nb};  e)Evolution of the Gaussian peak location normalized to the loop length ($s\peak/l$), controlled by parameter $q_2$ in Equation~\ref{nb}; f) Evolution of the flaring fluxtube radius $R$ at the reference point $s_0/l=-0.024$ (Equation~\ref{nb}), normalized by the extrapolation grid size, $dr=1.45\times10^8$~cm; the dotted line shows the corresponding linear fit; g) Evolution of the $\delta_1$ and $\delta_2$ spectral indices of the double power-law nonthermal electron  distribution; h) Standard deviations of the synthesized microwave spectra relative to the microwave spectra observed by OVSA. The median standard deviation of 12.4\% is shown by the horizontal dotted line.}
\end{figure}

\begin{figure}
    \centering
    \includegraphics[width=0.96\columnwidth,clip]{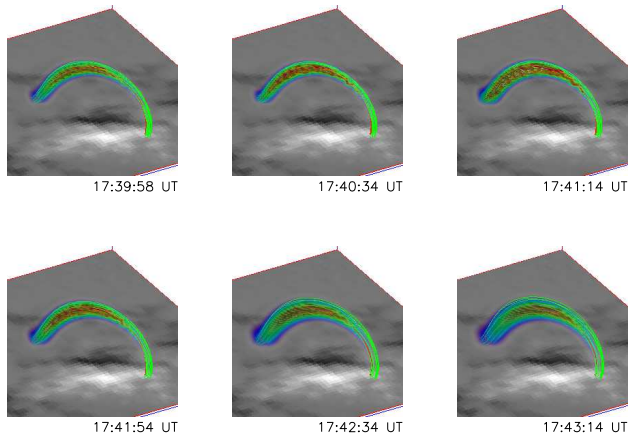}
    \caption{\label{fig_3d20020412_rise} Rise Phase of the SOL2002-04-12T17:42 flare: Evolution of the number density of the nonthermal electron distributions (shown by the color hue in the same scale throughout all panels in the figure) along the modeled flux tube in 3D. (An animation of this figure is available.)}
\end{figure}

The spatial distribution of the nonthermal electrons along the loop is described by a gaussian function

\begin{equation}
\label{nb}
n_{nth}(s)=n_b\exp\Big\{-\Big[q_0\Big(\frac{s-s_0}{l}+q_2\Big)\Big]^2\Big\},
\end{equation}
where $l$ represents the length of the loop central field line,  the parameter $q_0$ controls the width of the gaussian, and $q_2$ indicates where the gaussian has a maximum relative to a pre-selected $s_0$ reference point (e.g., the loop-top)\footnote{The adopted convention is that the coordinate $s$ along the central (reference) field line of the flaring flux tube is zero where the magnetic field is minimal by the absolute value, $s=s_{\min}<0$ at the footpoint of the positive magnetic field, while $s=s_{\max}>0$ at the footpoint of the negative magnetic field; $s_{\max}-s_{\min} = l$, where $l$ is the field line length.}.

\begin{figure}
    \centering
    \includegraphics[width=0.96\columnwidth,clip]{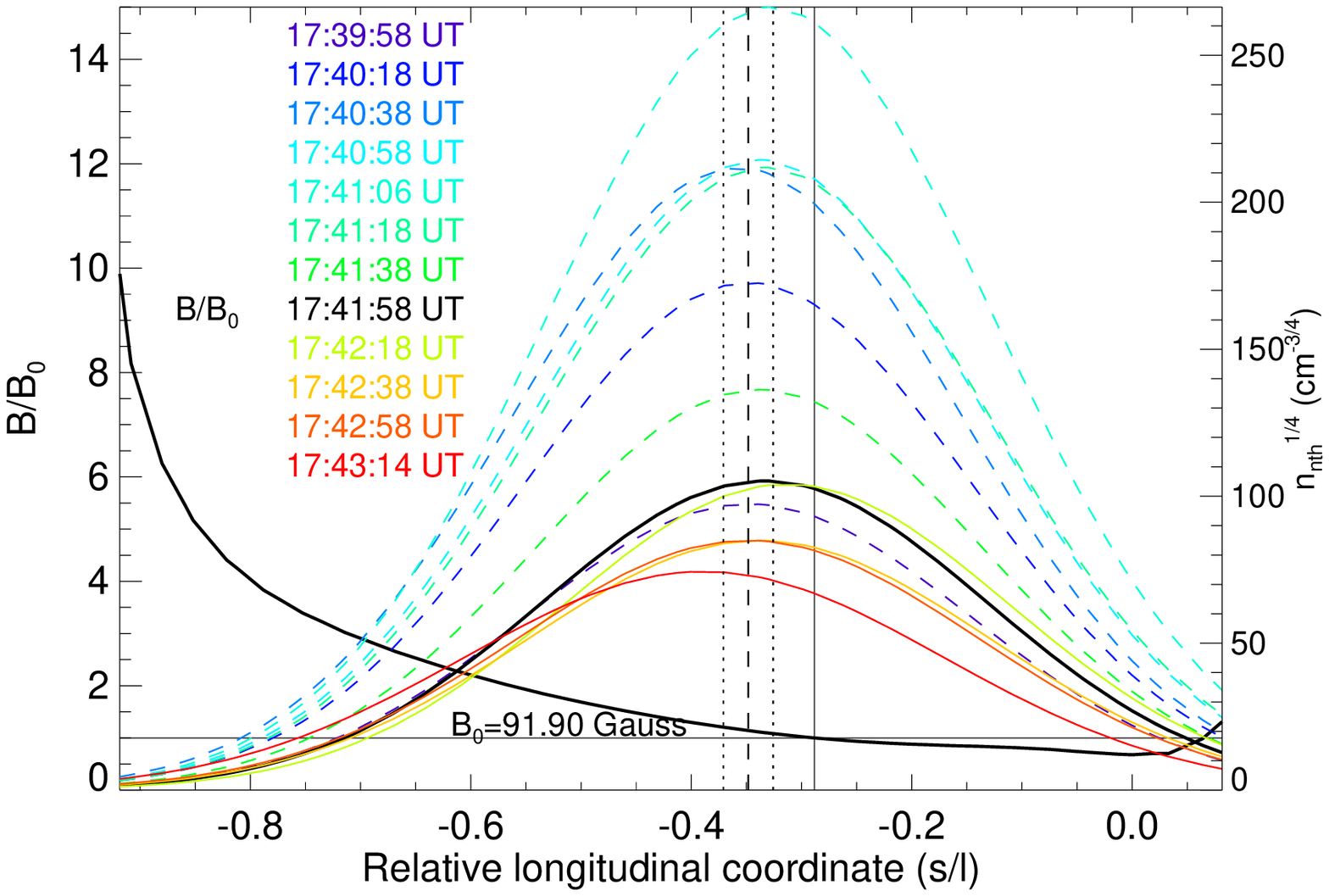}\\
    \caption{\label{nb20020412rise} Rise Phase of the SOL2002-04-12T17:42 flare: Variation of the absolute magnetic field along the central field line (left-hand vertical axis) and evolution of the longitudinal nonthermal electron distribution (right-hand vertical axis) described by Equation~\ref{nb}. The horizontal axis, $(s/l)$, indicates the signed distance from the loop apex $(s\equiv0)$ normalized to the total length of the central field line, $l=5.7\times10^9$~cm.  The absolute magnetic field line (black solid line) was normalized to the reference value corresponding to $s_0/l=-0.28$ shown by the vertical solid line. 
    For easier direct comparison, the Gaussian longitudinal density distributions $n_{nth}(s)$ were power-law scaled as $n_{nth}^{1/4}$. The evolution of the density distribution was color and line-style coded, the dashed and solid lines, respectively, representing time frames before and after the local peak time, which occurred at 17:41:58 UT. For clarity, only a few selected time frames were displayed, including the local peak time (black solid curve) and the time at which the maximum density is reached, i.e. 17:41:06 UT (dashed light-blue curve). The vertical dashed line indicates the mean location of the density distribution peaks, $s\peak/l=-0.34$, and the dotted lines indicate the corresponding $\pm\sigma=0.04$ standard deviation interval.}
\end{figure}

Figure~\ref{parms20020412rise} displays the evolution of the adjusted parameters used in the modeling along with some derived parameters.
Figure~\ref{parms20020412rise}a,d,e displays  parameters of the nonthermal electron gaussian distribution along the central field line for the selected magnetic fluxtube, according to Equation~\ref{nb}. Panel (d) shows that the gaussian width stays almost constant for more than two minutes of the rise phase and then slightly drops (the electron distribution becomes slightly more uniform along the loop spine). As shown by panel (e), the location of the spatial peak of the distribution, controlled by $q_2$, varies within $s\peak/l=s_0/l-q_2\approx0.35\pm0.04$. It does not display any monotonic trend but rather relatively modest variation around a ``preferred'' location at the middle of the fluxtube.
To provide an overview of the evolution of the nonthermal electron spatial distribution {visualized in animated Figure~\ref{fig_3d20020412_rise}}, the evolution of Gaussian nonthermal electron distribution along the spine of the flaring fluxtube is illustrated at a few selected time frames in Figure~\ref{nb20020412rise}.

However, it turns out that such redistributions only cannot account the observed build up of the optically thick \mw\ emission at the rise phase; thus, we are forced to increase the source area by increasing the cross-sectional radius of the flaring flux tube. Figure~\ref{parms20020412rise}f shows an almost monotonic rise of this radius from 0.9 to 1.6 grid points, which for our model resolution of two arcseconds ($dr=1.45\times10^8$~cm), corresponds roughly to a range from 1300 to 2300 km, which implies an increase of the flaring loop volume by a factor of three.
The information from panels (a), (d), (e), and (f) is used to compute the total number {and total energy} of the nonthermal electrons $N_{nth}$ in the model fluxtube shown in panel (b). Then, combining the inputs of panels (a) and (b) we compute an effective fluxtube volume as $V_{\rm ft}=N_{nth}/n_{nth}$ shown in panel (c). Finally, panel (h) displays the evolution of the model-to-data standard deviation, which characterizes  goodness of the model.


The apparent spectral evolution of the optically thin \mw\ emission (softening and then hardening) during roughly first two minutes of our four minutes time interval requires a corresponding ``hard-soft-hard'' evolutionary pattern in the nonthermal electron spectrum. Given that the model spectrum is described by a double power-law with a break up at about 36~keV, the mentioned spectral evolution pertains to the higher-energy electrons (spectral index $\delta_2$), that produce the \mw\ emission. Figure~\ref{parms20020412rise}g shows that $\delta_2$ gets softer from 3.4 at the very beginning of the interval  to 5.5 and then comes back to 3.4; it stays at 3.4 the two remaining minutes of this interval. The low-energy spectral index, $\delta_1$ stays constant almost all the time, $\delta_1=7.3$, but slightly hardens to $\delta_1=7$ during a few time frames at the middle of the interval. Although this change does improve the match with the data, we do not think it is a significant trend, so we are not discussing it in any detail.

The number density of the nonthermal electrons (above 18~keV adopted to be the low-energy cutoff in the spectrum) shows a pattern correlated with the $\delta_2$ evolution during the first two minutes, which is expected: the softer spectrum implies much fewer high-energy electrons for the fixed $E_{\min}$ and $E_{\rm break}$; thus, to compensate for that given the comparable \mw\ flux level, one has to increase the nonthermal electron distribution $n_{nth}$. It is instructive, however, to take a closer look at the $n_{nth}$ (Figure~\ref{parms20020412rise}a) evolution during the last two minutes of the interval, when no spectral evolution is detected (both $\delta_1$ and $\delta_2$ are constant in time). In spite of the rise of the radio flux, the peak (over the source) number density goes down by a factor of two. This result, however, is not surprising given that the source volume and, thus, the source area,  goes up during the same time; thus, the optically thick emission slowly goes up, while the optically thin one stays roughly at the same level.

The quality of the forward-fitted models is quantified by the relative model-to-data standard deviations displayed in Figure~\ref{parms20020412rise}h, which range from 7--37\%, with a median value of  12\%. This value shows a systematic decrease (the quality of the fit increases) with time, which is a direct outcome of the rising signal-to-noise ratio as the burst level goes up.

\begin{figure}
    \centering
    \includegraphics[width=0.96\columnwidth,clip]{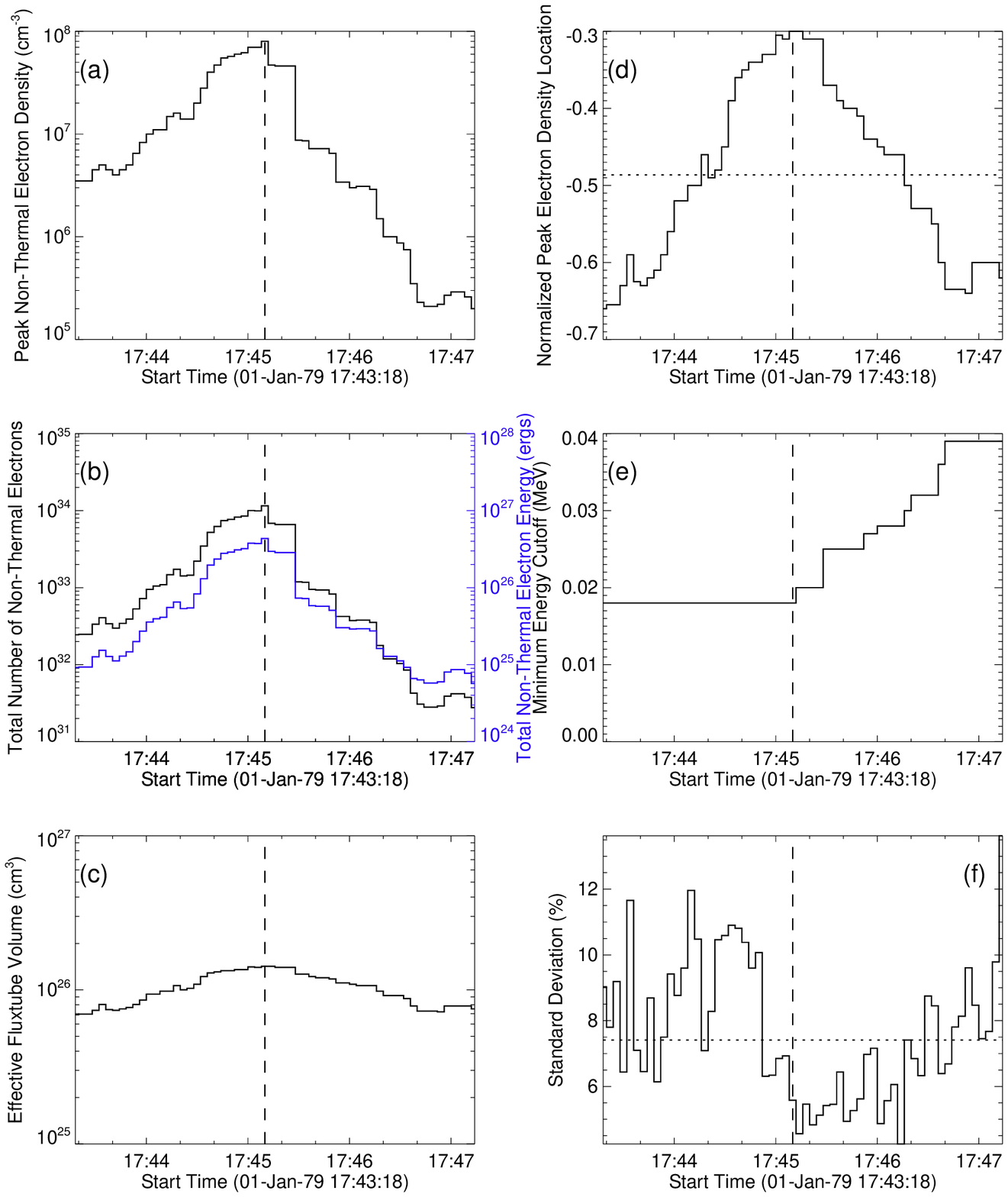}\\
    \caption{\label{parms20020412peak} Peak Phase of the SOL2002-04-12T17:42 flare: Evolution of the modeled parameters describing the spatial and energy nonthermal electron distributions around the main peak time 17:45:10~UT, which is marked by vertical dashed lines in each panel. a) Evolution of the Gaussian distribution full maximum ($n_{nth}$ in Equation~\ref{nb}); b) total number $N_{nth}$ and total energy of the nonthermal electrons; c) effective volume of the fluxtube; d)Evolution of the Gaussian peak location normalized to the loop length ($s\peak/l$), controlled by parameter $q_2$ in Equation~\ref{nb}; e) Evolution of the minimum nonthermal electron energy cutoff; f) Standard deviations of the synthesized microwave spectra relative to the microwave spectra observed by OVSA. The median standard deviation of 7.8\% is shown by the horizontal dotted line.}
\end{figure}

\subsection{Peak Phase of SOL2002-04-12T17:42}
\label{S_peak_20020412}

For the peak time, $\sim$17:45:10~UT, we used the \citet{Fl_Xu_etal_2016} master model without any adjustment and advanced it backward in time by 88 sec and forward by 2 minutes. It is interesting that adjusting only two parameters, location of the spatial peak of the nonthermal electron distribution along the loop spine, $q_2$, and the number density is sufficient to successfully model the emission rise phase just prior to the spectral peak. In the decay phase, just after the peak, the same parameters evolve as well, but adjusting one more parameter, namely $E_{\min}$, which affects the \mw\ spectral shape at the optically thick part of the spectrum, improves the model-to-data match remarkably{, see Figure~\ref{parms20020412peak}. This Figure also shows a few derived parameters such as total number of the nonthermal electrons, their total energy, and the effective volume of the flaring flux tube.}

\begin{figure}
    \centering
    \includegraphics[width=0.96\columnwidth,clip]{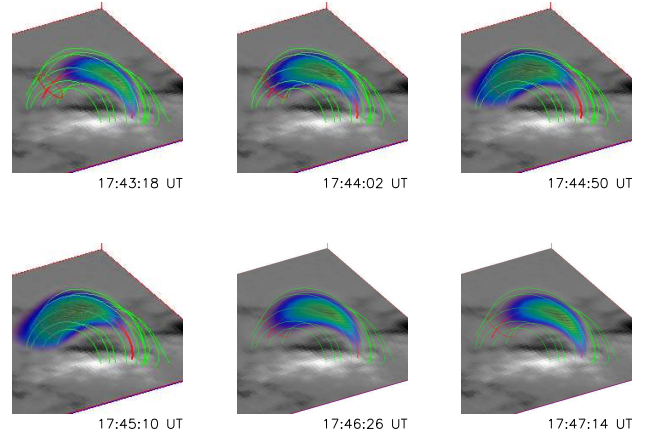}
    \caption{\label{fig_3d20020412_peak} Peak Phase of the SOL2002-04-12T17:42 flare: Evolution of the number density of the nonthermal electron distributions (shown by the color hue in the same scale throughout all panels in the figure) along the modeled flux tube in 3D. (An animation of this figure is available.)}
\end{figure}

\begin{figure}
    \centering
    \includegraphics[width=0.9\columnwidth,clip]{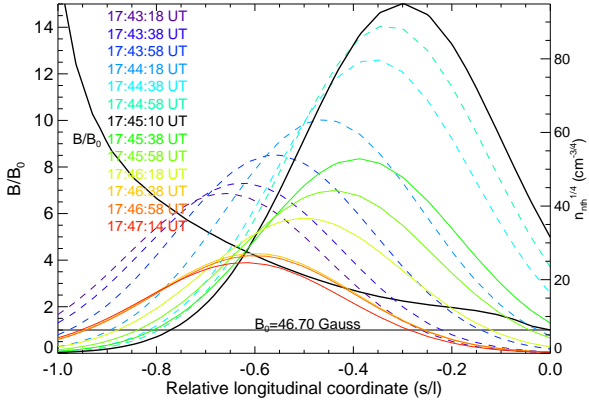}\\
    \caption{\label{nb20020412peak} Peak Phase of the SOL2002-04-12T17:42 flare: Variation of the absolute magnetic field along the central field line (left-hand vertical axis) and evolution of the longitudinal nonthermal electron distribution (right-hand vertical axis) described by Equation~\ref{nb}. The horizontal axis, $(s/l)$, indicates the signed distance from the loop apex $(s\equiv0)$ normalized to the total length of the central field line, $l=4.77\times10^9$~cm.  The absolute magnetic field line (black solid line) was normalized to the minimum value corresponding to the loop apex, $B_0=46.70$~Gauss. For easier direct comparison, the Gaussian longitudinal density distributions $n_{nth}(s)$ were power-law scaled as $n_{nth}^{1/4}$. The evolution of the density distribution was color and line-style coded, the dashed and solid lines, respectively, representing time frames before and after the peak time, which occurred at 17:45:10.}
\end{figure}

Interestingly, over the one and a half minutes of the rise phase, the number density increases by a factor of 20, from $4\times10^6$ to $8\times10^7$~cm$^{-3}$, while the peak \mw\ flux rises by only a factor of three. This happens because the nonthermal electron cloud moves gradually toward a flux tube region having a weaker magnetic field, {see animated Figure~\ref{fig_3d20020412_peak} and Figure~\ref{nb20020412peak}.} At the decay phase, the parameter trends revert. However, in addition to these trends, the value of the low-energy cutoff, $E_{\min}$, goes up roughly from 20 to 40~keV at the decay phase. Other than that, no significant spectral evolution is demanded by the data.

\section{Multiloop Flare}

The M6.5 class SOL2015-06-22T17:50 flare \citep{2016NatSR...624319J, 2016NatCo...713104L, 2017NatAs...1E..85W} was well observed by a new generation of high-resolution instruments including Goode Solar Telescope at BBSO and Expanded OVSA (EOVSA; Gary et al., in preparation). This flare represents an example contrasting to the dense SOL2002-04-12T17:42 flare given that it contained a relatively large flux tube dominating low-frequency radio emission similar to a few other recently reported flares \citep{Fl_etal_2016, Fl_etal_2017}.

\begin{figure}
    \centering
    \includegraphics[width=0.96\columnwidth,clip]{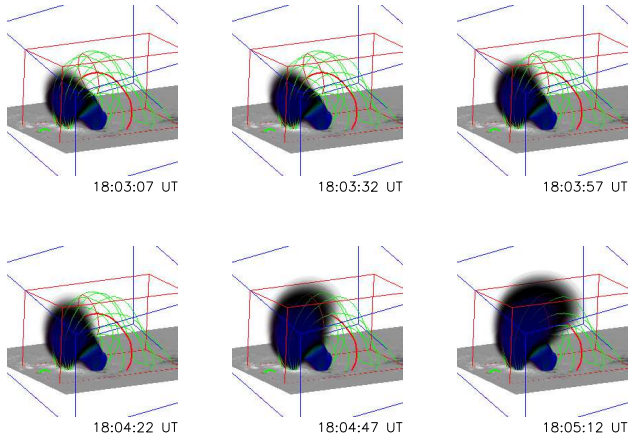}
    \caption{\label{3d20150622} SOL2015-06-22T17:50 flare: Evolution of the number density of the nonthermal electron distributions (shown by the color hue in the same scale throughout all panels in the figure) along the modeled flux tube in 3D. The large ``overarching'' loop is Loop~4, and the one beneath it is Loop~1. (An animation of this figure is available.)}
\end{figure}

\subsection{Overview of SOL2015-06-22T17:50}

\citet{2018ApJ...852...32K} performed a detailed 3D modeling of a single time frame corresponding to a local peak, 18:05:32 UT, of this flare using a magnetic data cube obtained from a nonlinear force-free field (NLFFF) reconstruction as a framework. Using this data cube, the flux tubes were identified needed to fulfill all constraints available from X-ray (\rhessi) and \mw\ (EOVSA) observations. It is interesting that three distinct flux tubes were needed to reproduce a rather complicated spatial distribution of X-ray images at different energies. However, even with these three loops it was impossible to reproduce the \mw\ low-frequency emission; neither the spectrum, nor the interferometric data. This forced \citet{2018ApJ...852...32K} to add one more loop, a large ``overarching'' loop, making the main contribution to the \mw\ spectrum. Presumably, this large loop was magnetically connected with other flaring loop(s) and served as a particle trap.

\begin{figure}
    \centering
    \includegraphics[width=0.98\columnwidth,clip]{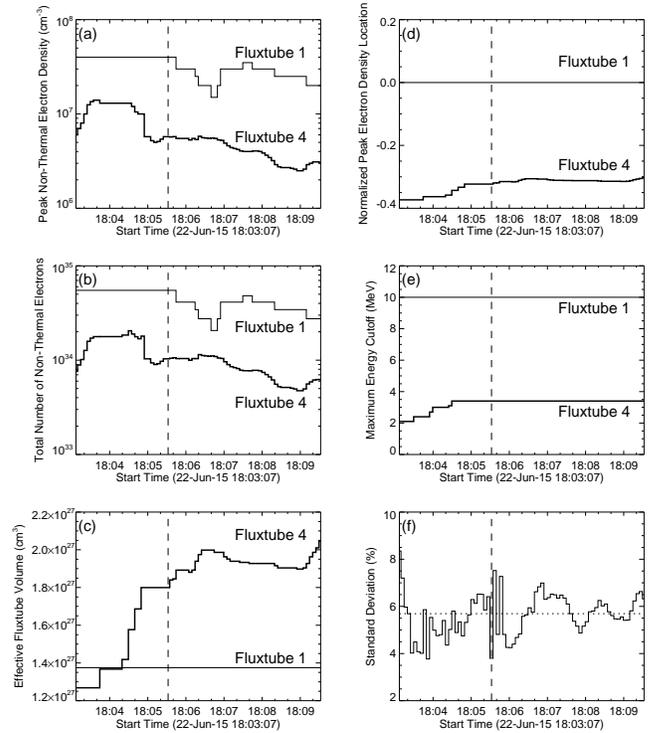}\\
    \caption{\label{parms20150622} SOL2015-06-22T17:50 flare: Evolution of the modeled parameters describing the spatial and energy nonthermal electron distributions around the main peak time 18:05:32~UT, which is marked by vertical dashed lines in each panel. a) Evolution of the Gaussian distribution full maximum ($n_{nth}$ in Equation~\ref{nb}); b)  total number of nonthermal electrons $N_{nth}$; c) effective volume of the fluxtube; d) Evolution of the Gaussian peak location normalized to the loop length ($s\peak/l$), controlled by parameter $q_2$ in Equation~\ref{nb}; e) Evolution of the maximum nonthermal electron energy cutoff; f) Standard deviations of the synthesized microwave spectra relative to the microwave spectra observed by EOVSA. The median standard deviation of 5.7\% is shown by the horizontal dotted line.}
\end{figure}

\begin{figure*}
    \centering
    \includegraphics[width=0.45\textwidth,clip]{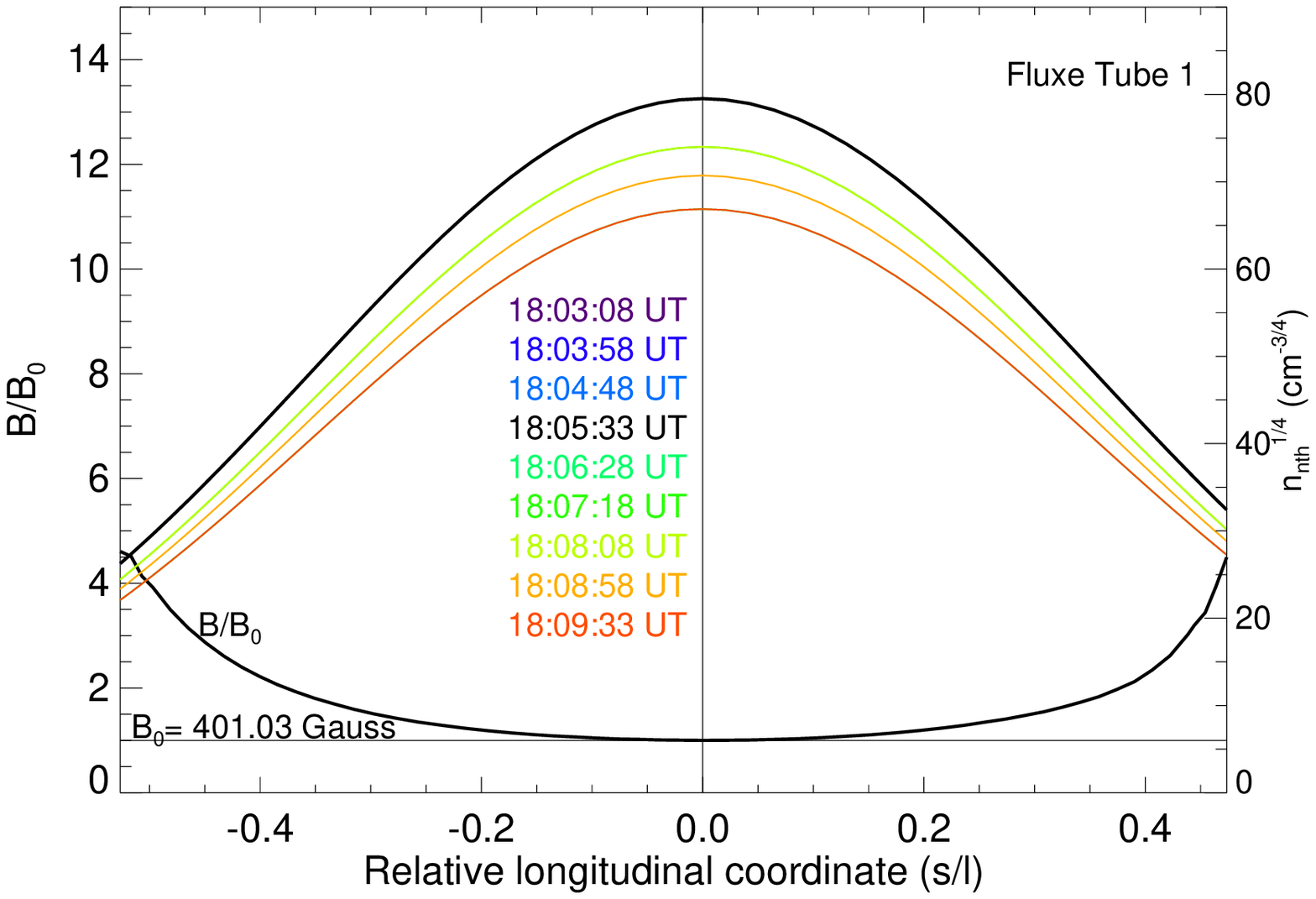}
    \includegraphics[width=0.45\textwidth,clip]{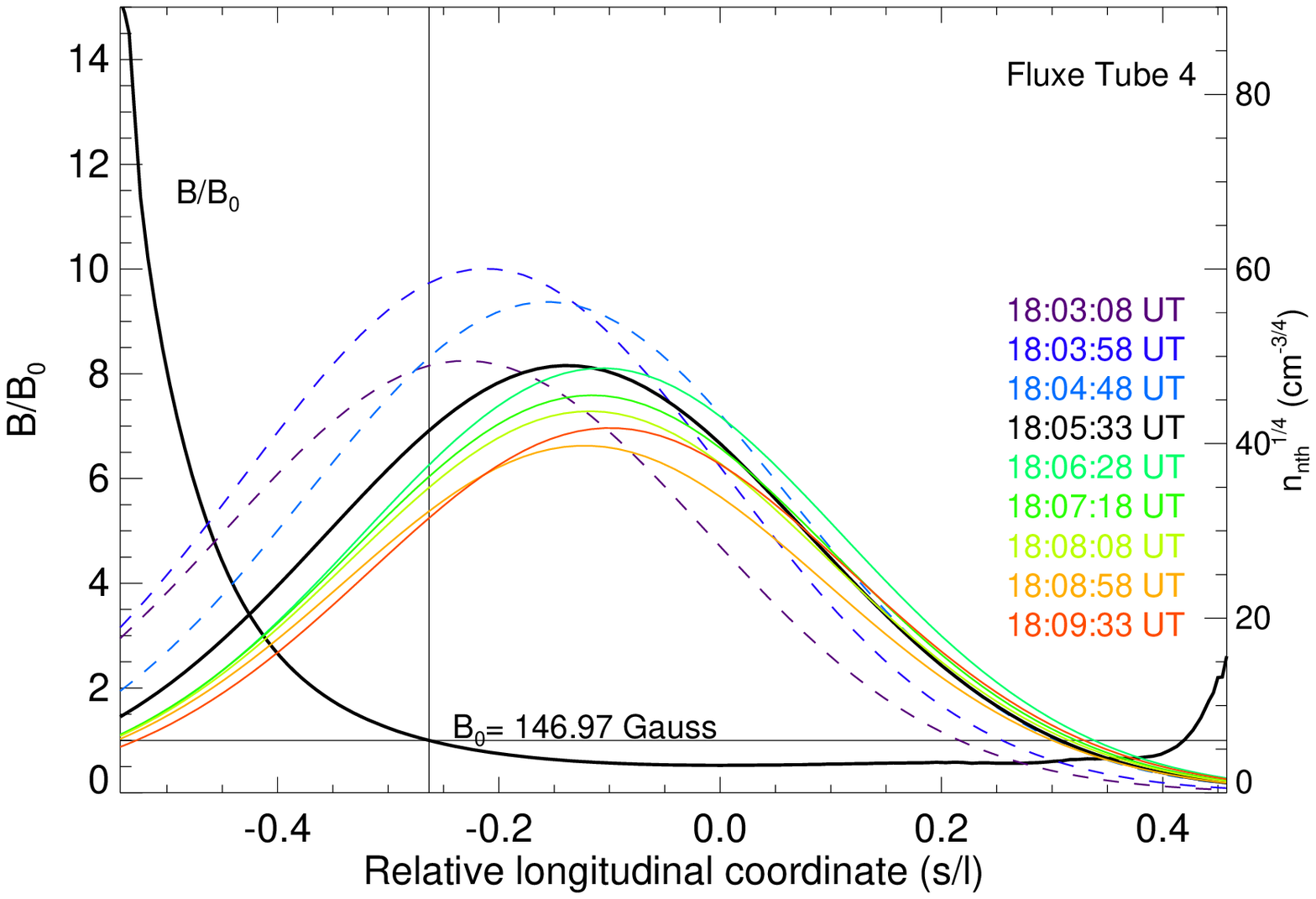}\\
    \caption{\label{nb20150622} SOL2015-06-22T17:50 flare: Evolution of the longitudinal nonthermal electron distributions along two of the four modeled flux tubes.}
\end{figure*}

It is important to note that NLFFF reconstruction ensures a measurably better approximation to the reality than other modeling means such as linear force-free or potential field  reconstruction; thus, for the first time, it became possible to build those various flaring loops using one single data cube, rather than employing various cubes as in previous studies \citep[see, e.g.,][]{Fl_etal_2016, Fl_etal_2017}.
However, given that the NLFFF data cube is only an approximation to the reality, which contains noticeable uncertainties \citep{deRosa_etal_2009, 2017ApJ...839...30F}, these uncertainties propagate to the final model and do not permit a perfect match of all observables. Specifically, \citet{2018ApJ...852...32K} had to trade off between reproducing the \mw\ spectrum and relative visibilities (interferometric spatial information; see Kuroda et al. 2018 for more detail). To match the \mw\ spectrum, two distinct sources were needed, while the relative visibilities suggested that there was either only one single source or the two sources must have projected at the same place. \citet{2018ApJ...852...32K} managed to obtain apparently correct spatial structure of two sources that matched the relative visibility constraints at the expense of noticeable spectral distortion.
We believe that this close spatial association between the two \mw\ producing loops is not a random coincidence but rather comes from a connectivity between these loops, which is, however, not captured by the static NLFFF model, which cannot be fully correct especially at the areas of loop interaction during the flare. Thus, having certain mismatches between model and data is not surprising.
Given that in this study we are guided by solely the evolving \mw\ spectral shape, we modified the original \citet{2018ApJ...852...32K} model such as to get the best possible spectral match at 18:05:32 UT. Although the modification is relatively minor, the new model contains slight spatial displacement between the two \mw\ sources, which results in a measurably worse model relative visibilities.

\subsection{Peak phase of SOL2015-06-22T17:50}

Starting with this modified master model at 18:05:32 UT, we investigated 6.5 minutes of the flare evolution, 2.5 minutes before and four minutes after the master time frame with 5~sec cadence (the data are available with a higher, 1~s, cadence, but we did not see any fast variation of the data, which would warrant that high cadence in the modeling). Although two of the four flaring flux tube (Loop 1 and 4; see Fig.~\ref{3d20150622}) contribute to the \mw\ emission, only parameters of the overarching loop (Loop 4) show significant evolution, while in the other loop (Loop 1) only one parameter, the number density of the nonthermal electrons, evolves, {see Figure~\ref{parms20150622}. We do not show the total electron energy in this case given that the value of the low-energy cut-off is not well constrained in the large loop; on top of that, this loop is only one of four flaring loops, which are present in this flare, so the information about total nonthermal energy evolution in this flare cannot be obtained from our modeling.} Similarly to the peak time of the SOL2002-04-12T17:42 flare, the evolutionary patterns at the rise and decay phases are different from each other, which is not surprising.

At the rise phase, the modeling, {visualized in animated Figure~\ref{3d20150622} and further quantified in Figure~\ref{nb20150622},} shows a significant spatial evolution of the nonthermal electron population in Loop 4, which can be interpreted as a gradual filling of this loop by nonthermal electrons coming from another place, presumably, the acceleration region of the flare. The number density of nonthermal electrons in Loop 4 increases at the beginning of the rise phase, but then stays relatively constant and even starts to go down before the peak time. However, given the motion of the nonthermal electron cloud towards the loop top which has bigger cross-sectional area, the 
effective area of the \mw\ source associated with
Loop 4 continues to grow. It is also interesting that the high-energy cutoff in the electron spectrum grows at the beginning of the rise phase from 2.1 MeV up to 3.4 MeV and then stays constant. This may indicate that the acceleration of electrons up to MeV energies takes longer than for HXR-producing deka-keV electrons. We cannot, however, distinguish if this acceleration happens at Loop 4 or these high-energy electrons arrive there from the acceleration region.

In contrast, in the decay phase we almost do not see any spatial evolution of the nonthermal electron population and no spectral evolution, but rather gradual decrease of the number density (and, thus, total number) of nonthermal electrons in Loop 4, which we interpret as a gradual loss of the trapped electrons from this loop, while the supply from the nonthermal electron source gets weaker. This is consistent with that the number density in Loop 1 also goes down, while showing some variations at the decay phase. This consideration shows that in contrast to significant spatial complexity, this flare displays a relatively simple time evolution.

\section{Discussion}
\label{S_Discuss}

{In this section we  compare the identified properties of these two contrasting flares and use the revealed quantitative parameter trends to constrain  possible mechanisms of electron acceleration and regimes of their transport.}

\subsection{Comparison between Dense and Tenuous Flaring Flux Tubes}

The flare SOL2002-04-12T17:42 displayed only one flux tube in both X-rays and \mw s. The thermal density of this loop proved to be rather high, $\sim10^{11}$~cm$^{-3}$. This estimate comes from  analysis of the X-ray source sizes at different energies \citep{Xu_etal_2008}, from {the} emission measure determined from the X-ray spectral fit, and {it is} also consistent with joint 3D X-ray and \mw\ modeling performed by \citet{Fl_Xu_etal_2016}. In addition, the high plasma density is also supported by the presence of a distinct low-frequency radio component co-spatial (at least, in the projection) with the flaring loop seen in X-rays and \mw s, see Figure~\ref{f_20020412_images}. This low-frequency component is likely produced by RTR \citep[cf.][]{Nita_etal_2005} around the plasma frequency; having the plasma frequency around 2~GHz implies the electron number density of about $5\times10^{10}$~cm$^{-3}$. In that high plasma density, the Coulomb loss time is very short, $\lesssim1$~sec, for the deka-keV electrons responsible for HXR emission. This means that the electrons must be energized inside the loop; thus, this dense loop represents (or inscribes) the acceleration region of this coronal thick-target flare. This conclusion is confirmed by the source modeling performed by \citet{Xu_etal_2008}. Therefore, the trends of the flare parameters revealed in Section~\ref{S_dense_flare} pertain, in fact, to evolution of the acceleration region in this flare. In contrast, most of the dynamics revealed in the case of the SOL2015-06-22T17:50 flare pertains to the large, ``overarching'' flux tube; thus, in this case we primarily study the evolution of the electron trapping, rather than {the} acceleration. From this perspective, and also given highly different sizes of the flux tubes and the corresponding number densities, it is not at all surprising that these two flares demonstrate different evolutionary patterns.

\subsection{Dynamics of Electron Acceleration in the Dense Flare}
At the early rise phase of the SOL2002-04-12T17:42 dense flare (Figures~\ref{parms20020412rise}--\ref{nb20020412rise}), the main finding is {related to} how the flare emission builds up as the flare progresses. Specifically, we found that the increase of the \mw\ flux happens because of {the} rising volume of the flaring loop, rather than the {evolution of the} nonthermal number density. The number density of the nonthermal electrons can even go down but the \mw\ source area  
goes up. so the low-frequncy optically thick radio flux goes up proportionally.
It is reasonable to assume that the increase of the flaring loop volume occurs because of the accumulation of the newly reconnected magnetic flux tubes. Likely, more non-potential magnetic subdomains become unstable earlier, and then trigger adjacent, less non-potential subdomains. This might explain why the number density of the accelerated electrons goes down during this process: less free energy density implies lower number density of the accelerated electrons.
Some spatial and spectral evolution of the nonthermal electron population occurs during this build up; however, this evolution is modest. 
The spectral evolution is over by the end of this early rise phase.


A relatively slow early rise of the radio flux gives a way to a much faster rise at about 17:43:42~UT, when {the} evolutionary pattern significantly alters compared with that at the early rise phase. Indeed, no change of the flaring loop topology or size is demanded by the data at this stage (recall, that the master models for these two time intervals are different because of some topology change over the course of the flare), while the rise of the flaring emission between 17:43:42~UT and the peak at 17:45:10~UT is solely due to increase of the nonthermal electron number density by a factor of 20. However, this { number density} increase resulted in only a factor of 3--4 increase of the \mw\ flux, because the newly accelerated electrons occupied area with accordingly weaker magnetic field. The fact that no significant topological changes happens at this stage implies that the flaring loop itself contains a sufficient amount of free energy to support {the} acceleration of electrons.

After the peak at 17:45:10~UT, the trends in the parameter behavior revert: the number density of the nonthermal electrons goes down, while the peak of their spatial distribution returns to the portion of the loop with a larger magnetic field. In addition, the low-energy cutoff in the electron energy spectrum goes up from roughly 20~keV to 40~keV, over {a two minute interval in} this decay phase. A naive interpretation of this trend as due to the collisional loss does not work because the collisional loss time is about 1~sec or less at this energy range. We instead propose that this behavior indicates a change in the balance between the acceleration and losses, as follows. At the rise phase, a yet unspecified acceleration agent has sufficient energy to overcome the Coulomb losses at essentially all nonthermal energies, which leads to the growth of the nonthermal population. At some point, however, 
the free energy  available for particle acceleration goes down. Thus, the acceleration efficiency goes down such as it cannot overcome the Coulomb losses at low energies any longer, but {it is} still sufficient to compensate the Coulomb losses of the electrons above some threshold energy. The lower the acceleration efficiency, the higher  the threshold energy below which the acceleration is  no longer efficient.

\subsection{Casting of Acceleration Mechanisms in the Dense Flaring Flux Tube}

Let us consider the corresponding implications for the acceleration mechanism. {The main} groups of the acceleration mechanisms are acceleration by (i) DC electric field, (ii) resonant turbulence, and (iii) non-resonant turbulence. Acceleration due to DC electric fields is known to be efficient at relatively low energies, say $\lesssim 100$~keV \citep{1985ApJ...293..584H}. However, electrons with much higher energies are involved in generation of the \mw\ emission of the flare; thus, we will concentrate here on stochastic electron acceleration by turbulence. It is reasonable to expect that the acceleration efficiency is higher in sub-volumes with accordingly higher free energy; thus, in locations with a relatively strong magnetic field. For this reason, in our estimates we place the center of the acceleration region at $s/l\approx-0.8$, where $B=7\times B_0\approx327$~G given that $B_0=46.7$~G. This is consistent with the location of \rhessi\ X-ray source interpreted as the acceleration region by \citet{Xu_etal_2008, 2012A&A...543A..53G, 2012ApJ...755...32G, 2013ApJ...766...28G}.

\begin{figure*}
    \centering
    \includegraphics[width=0.9\textwidth,clip]{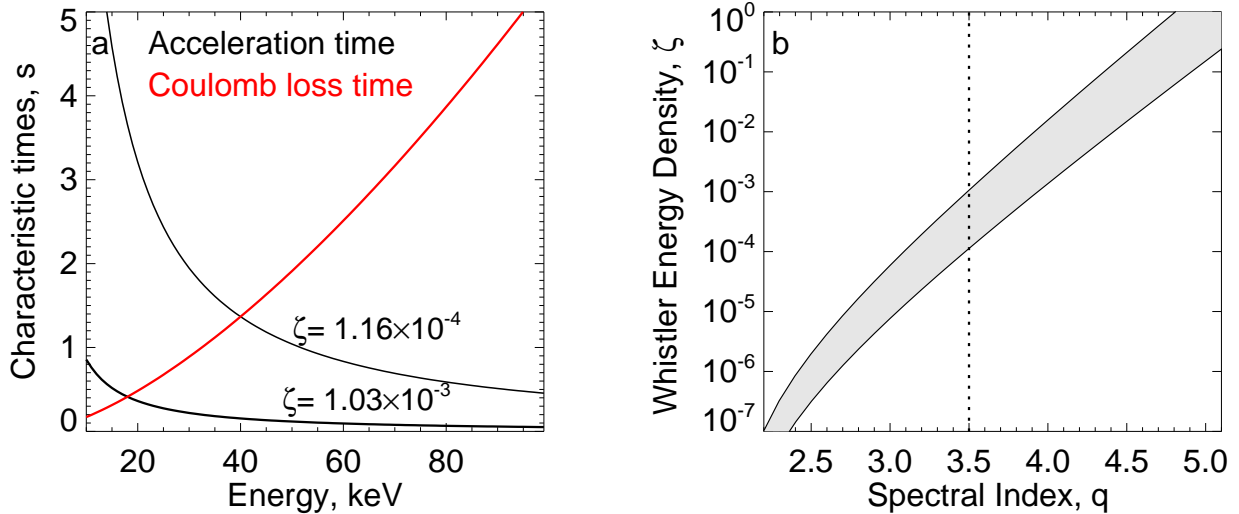}\\
    \caption{\label{f_acc_times_whistlers} (a) Acceleration time as a function of electron energy computed for whistler turbulence with spectral index $q=3.5$ (shown by a dashed line in panel b) and two different levels of the energy density $\zeta=W_{\rm whist}/W_B$ shown in the panel selected such as to match the mean Coulomb loss time for two energies, 18~keV and 40~keV, in our modeling flux tube with $\left<n_0\right>=3.7\times10^{10}$~cm$^{-3}$ \citep[see Table 2 in][]{Fl_Xu_etal_2016}. (b) The range of the whistler turbulence energy density needed to compensate the Coulomb losses for electron energies between 18 and 40~keV for various spectral indices of the whistler turbulence.}
\end{figure*}

\subsubsection{Nonresonant Acceleration}

Stochastic acceleration by a large-scale, non-resonance turbulence requires that the nonthermal particles are linked to a ``fluid element'' of the plasma due to efficient particle scattering by a small-scale, resonant turbulence \citep[][\S 11.5]{Byk_Fl_2009, Fl_Topt_2013_CED}, such as the local meam free path (mfp) $\lambda$ of the nonthermal electrons, controlled by electron scattering by resonant turbulence, is much shorter than the main, energy-containing scale $L_c$; $\lambda<L_c$. In such a case, the particles are being pumped up with the energy of the large-scale turbulent pulsations. Given that the source of the free energy is the nonpotential portion of the magnetic field, the turbulent plasma velocity $u$ cannot exceed the Alfven velocity $v_A$. The characteristic acceleration time, which can, in the general case, be estimated as the inverse diffusion coefficient in the energy space, $\tau_{acc}\approx 1/D(E)$, for the nonresonant stochastic acceleration is $\tau_{nonres} \sim (9L_c/v_A)(v_A/u)^2 \geq 9L_c/v_A$, and it is independent of the nonthermal electron energy \citep[][Eq. 3]{Byk_Fl_2009}. For the selected value of the magnetic field $B\approx327$~G and the mean thermal number density $\left<n_0\right>=3.7\times10^{10}$~cm$^{-3}$ \citep[see Table 2 in][]{Fl_Xu_etal_2016} known from the model, the Alfven speed is $v_A=3.7\times10^8$~cm/sec. To compensate the Coulomb losses of the electrons at 18~keV, the acceleration time must be shorter than the Coulomb loss time at that energy, $\tau_{loss}\approx0.4$~sec \citep[see, e.g., Eq. 4 in][]{bastian_etal_2007}, which requires  $L_c\leq v_A \tau_{loss}/9\approx 1.67\times10^7$~cm; the equality corresponds to $u\approx v_A$, while for $u< v_A$ the turbulence main scale has to be proportionally smaller. At 40~keV, $\tau_{loss}\approx1.37$~sec; thus the large-scale, nonresonant turbulence kinetic energy needed to compensate the Coulomb losses at 40~keV is a 30\% fraction of that needed to compensate the Coulomb losses at 18 keV.

As {it} has been said, the applicability of the nonresonant stochastic acceleration regime requires that the local mfp $\lambda$ of the nonthermal electrons controlled by resonant scattering by small-scale turbulence is much shorter than $L_c$; $\lambda<L_c$. However, the required $L_c$ is already very short; thus, having an even shorter local mfp requires a rather high level of the resonant turbulence, which itself can effectively accelerate electrons. We note also that having that short mfp, $\lambda<10^7$~cm, may be in conflict with the observed sizes of the X-ray source, which is likely controlled by collisional transport \citep{Xu_etal_2008}, rather than wave-particle interaction. These considerations do not favor the nonresonant stochastic acceleration in the given event; thus, we now focus on the resonant stochastic acceleration by small-scale turbulence.

\subsubsection{Resonant Acceleration}

To be specific, here we consider stochastic acceleration by whistler turbulence following \citet[][\S~11.2.1]{hamilton_petrosian_1992, Fl_Topt_2013_CED}. In this case, the diffusion coefficient $D(E)$ and, thus, the acceleration time, depend on the electron energy, $D(E)\propto\beta(\beta\gamma)^{q-2}$, where $\beta=v/c$ the particle velocity normalized by the speed of light, $\gamma$ the Lorentz-factor, $q$ the spectral index of the resonant turbulence. For our estimates we adopt that the whistler spectrum is present between the proton and electron gyrofrequencies, $\omega_{Bp}$ and $\omega_{Be}$; thus, the main scale is $L_c\sim v_A/\omega_{Bp}\approx120$~cm, while its energy density $W_w$ is a fraction $\zeta$ of the magnetic field energy density, $W_w=\zeta B^2/(8\pi)$.   Figure~\ref{f_acc_times_whistlers}a demonstrates dependence of the acceleration time on the electron energy for $q=3.5$ and two different levels $\zeta$ of the whistler energy density selected such as the acceleration time to exactly match the Coulomb loss time at 18 and 40~keV, respectively. This plot shows that the evolution of $E_{\min}$ from 18 to 40~keV implies that the whistler turbulence decays by a factor of 9 (to be compared with a factor of 3 determined above for the nonresonant stochastic acceleration). The electron mfp due to scattering by the whistler waves $\lambda(18$~\textrm{keV}$)=4.9\times10^8$~cm for $\zeta=10^{-3}$ and $\lambda(18$~\textrm{keV})$=4.3\times10^9$~cm for $\zeta=1.16\times10^{-4}$ are comparable to the collisional loss length $\lambda_C(18$~\textrm{keV}$)=3.2\times10^9$~cm, in agreement with the currently established understanding of the coronal thick-target flares.

Figure~\ref{f_acc_times_whistlers}b displays the corresponding range of the whistler turbulence energy density as a function of the turbulence spectral index $q$. Strong dependence of the needed energy density on the spectral index simply follows from the fact that the electrons are accelerated by resonant waves whose wavelength is about the electron Larmor radius $R_0\sim1-2$~cm for the energy range of interest, while the energy density at these scales is a $(R_0/L_c)^{q-1}$ fraction of the total energy density $W_w$. Note, that the width of the shaded area in Figure~\ref{f_acc_times_whistlers}b corresponds to roughly one order of magnitude in $\zeta$ variation regardless the spectral index value.

\subsubsection{A Consistency Check}

\begin{figure}
    \centering
    \includegraphics[width=0.96\columnwidth,clip]{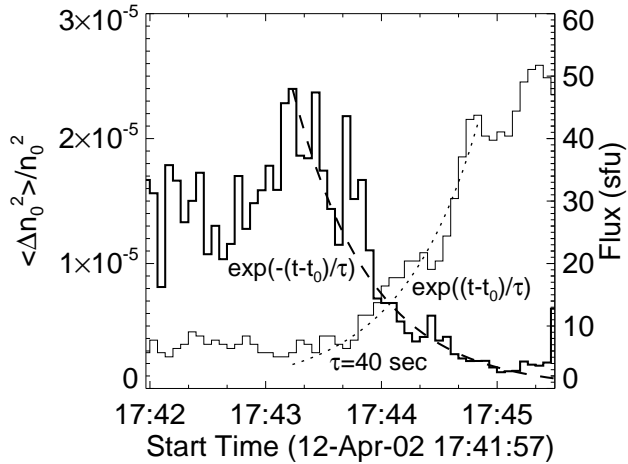}\\
    \caption{\label{f_dN2N2} Evolution of turbulent small-scale density inhomogeneities derived from the evolution of the radio flux at 1.4~GHz interpreted as the RTR produced by interaction of available nonthermal electrons with the density inhomogeneities. Parameters of the nonthermal distribution are taken from the evolving 3D model; see Figure~\ref{parms20020412peak}. The dashed curve that starts at the peak of the plot approximate the decay phase of the density inhomogeneity evolution by an exponential function with the characteristic time $\tau=40$~sec. 
    }
\end{figure}

Now, recall that the whistler mode is a high-frequency continuation of the fast mode \citep[see, e.g., Chapter~3 in][]{Fl_Topt_2013_CED}, which contains fluctuations of the plasma density along with fluctuations of the magnetic and electric fields. Interaction of the nonthermal electrons with such density fluctuations results in generation of a so-called transition radiation \citep[TR,][]{Pl_Fl_2002}. Resonant TR (RTR) arising at frequencies close to the plasma frequency has been reported in a number of solar flares \citep{2003ApJ...585..524L, RTR, Nita_etal_2005}. In Section~\ref{S_overview_20020412}, we have already noted that the radio burst displays a distinct decimetric continuum component, which we are inclined to interpret as RTR. According to \citet{Pl_Fl_2002}, the functional form of the RTR intensity varies depending on relationship between the spectral index of the density fluctuation spectrum, $\nu$, and the (low-)energy index in the nonthermal electron spectrum, $\delta_1$. In our case, $\delta_1=6.3$ \citep[see Table 2 in][]{Fl_Xu_etal_2016}; thus, for all reasonable spectral indices $\nu<8.6$, the inequality $\nu<2\delta_1-4$ takes place and the RTR flux density is specified by Equation~(13) in \citet{RTR}:

\begin{equation}
\label{Eq_RTR_flux}
 F \approx 2 \cdot 10^8 {\Delta f_0 \over 1~\textrm{GHz}} {\langle\Delta
n_0^{2}\rangle \over  n_0^{2}} \left({v_0 \over c}\right)^{\nu+2} 
{\int n_{nth}(>E_{\min}) dV \over 10^{36}} \times $$$$
{c^2 \over v_T^2} \left({f \over \Delta
f_0}\right)^{2} \exp\left[-{(f-f_0)^2 \over \Delta f_0^2}\right]
\quad {\rm sfu},
\end{equation}
where $\langle\Delta n_0^{2}\rangle$ is the mean square of the
small-scale density inhomogeneities at the scales $l<l_c=c/f_p\approx17$~cm, $n_0$ is the mean thermal plasma number
density, $v_0(E_{\min})$ is the minimum velocity in the nonthermal
electron distribution that corresponds to the low-energy cutoff $E_{\min}$, $N_{nth}=\int n_{nth}(>E_{\min}) dV$ is the total, integrated over the volume $V$, number of nonthermal
electrons at the source above the low-energy cutoff $E_{\min}$, $f_0$ is the mean plasma frequency at the source, $\Delta f_0$ is the standard deviation of the plasma frequencies at the source that reflects the large-scale source non-uniformity, $v_T$ is the thermal velocity of plasma electrons.

Remarkably, we can solve Equation~(\ref{Eq_RTR_flux}) for $\langle\Delta n_0^{2}\rangle /  n_0^{2}$, given that all but one ($\nu$) input parameters of this equation are known from either {the} model or data directly. Specifically, from the model we know {the} evolution of the nonthermal electron population; 
$v_T$ is known from the plasma temperature $T=21$~MK determined from the X-ray spectral fit \citep[see Table 1 in][]{Fl_Xu_etal_2016}; while $f_0=1.4$~GHz and $\Delta f_0\approx 1$~GHz are determined as the peak frequency and spectral width of the decimetric component from the radio data. Then, the evolution of the flux density $F$ at $f_0=1.4$~GHz is also determined from the radio data. Putting all these inputs together and assuming a moderate value of $\nu=1.5$, we display in Figure~\ref{f_dN2N2}  the  recovered evolution of the plasma density fluctuations $\langle\Delta n_0^{2}\rangle /  n_0^{2}$ during the electron acceleration phase (i.e., before $E_{\min}$ has started to grow, given that afterwards, the results depend on exact shape of the nonthermal electron spectrum below $E_{\min}$, which is underconstrained by the data). This evolution derived from the decimetric component (which is distinct from the \mw\ spectral component) can be considered as a proxy for the evolution of the turbulence responsible for acceleration of electrons that produce the \mw\ emission. 

The evolution such as seen in Figure~\ref{f_dN2N2} is actually expected within the stochastic acceleration process. The  rise of the turbulence at around 17:42:30~UT is followed by {the} electron acceleration episode seen in Figure~\ref{parms20020412peak}a. Interestingly, the peak in the nonthermal electron number density is delayed, by roughly 1.5~minutes, compared with the peak {time} of the turbulence, which is easy to understand. Indeed, while the turbulence level goes up above a given threshold, the acceleration is a dominant process that shapes the nonthermal electron distribution. This is true not only at the turbulence level peak time, but also in some vicinity of this peak at both rise and decay phases. This means that while the turbulence decays after its peak time, its energy is spent to electron acceleration. The turbulence energy transferred to the nonthermal electron population is initially still large enough to support the electron acceleration. Later on, the turbulence continues to further decay, such as it cannot compensate the Coulomb losses at low energy, but it is still powerful enough to support higher energy electrons. This is confirmed by more or less steady level of high-frequency \mw\ emission controlled by the high-energy electrons, accompanied by progressive loss of low-energy electrons due to the Coulomb losses. Note, that the decay phase of the $\langle\Delta n_0^{2}\rangle /  n_0^{2}$ evolution in Figure~\ref{f_dN2N2} is consistent with the exponential decay with a time constant of 40~sec. 

We conclude that the resonant stochastic acceleration of nonthermal electrons by whistler turbulence in the dense flaring loop offers a consistent interpretation of the source parameter evolution recovered from the 3D modeling and also co-spatial decimetric component quantitatively interpreted here as produced by RTR emission mechanism due to interaction between nonthermal electrons and small-scale density fluctuations, likely generated by the same whistler turbulence.


\begin{figure*}
    \centering
    \includegraphics[width=0.9\textwidth,clip]{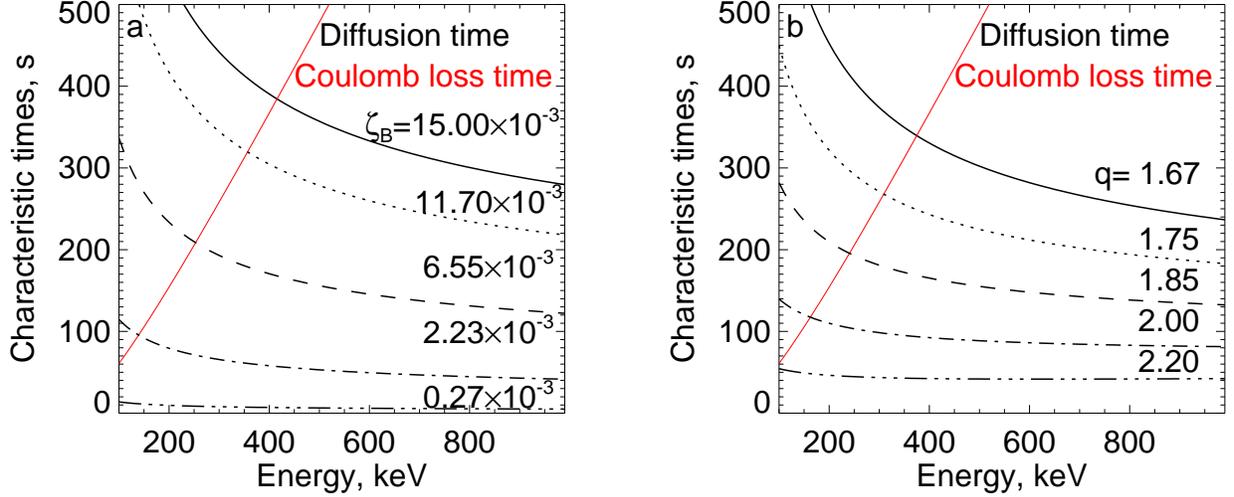}\\
    \caption{\label{f_diff_times} Diffusion time needed for an electron with a given energy to diffuse a length $4.5\times10^{9}$~cm equal to 30\% of the overarching loop length ($1.5\times10^{10}$~cm). (a) Diffusion time as a function of electron energy computed for a ``generic'' spectrum of magnetic inhomogeneities with the main scale $L_c=10^7$~cm, spectral index $\nu=1.67$ and the levels $\zeta_{\rm B}\equiv\left<\Delta B^2\right>/B^2$ indicated in the panel.
    (b) A similar plot but computed for the whistler turbulence spectrum with $\zeta=W_{\rm whist}/W_B=10^{-4}$, the main scale $L_c=v_A/\omega_{Bp}\sim 5$~m (corresponding to low-frequency cutoff of the whistler mode, which is still much larger than the Larmor radius of the nonthermal electrons, $\sim5-15$~cm), and various spectral indices $q$ indicated in the plot. Red lines indicate the Coulomb loss time.}
\end{figure*}

\subsection{Dynamics of Nonthermal Electron Trapping in a Large Tenuous Flux Tube}
In contrast, in the SOL2015-06-22T17:50 flare we mainly recover the dynamics of the trapping in a large loop: the trapped population of nonthermal electrons grows in space during the rise phase, while decays in number at the decay phase. Only the very beginning of the rise phase shows an indication of electron acceleration seen as an increase of the high-energy cutoff. It is unclear, however, if this acceleration takes place in the large loop itself, or in the loop interaction region, lower in the corona.

Let us consider the nonthermal electron evolution in the large loop. At the rise phase, the nonthermal electron cloud injected low at the eastern loop leg, presumably, from the region of the loop interaction, slowly fills the large loop up, over roughly two minutes of the flare evolution. Kuroda et al. (2018) showed that electrons with reasonably high energies, $\gtrsim300$~keV, are responsible for \mw\ emission from this loop, whose time of flight through the large loop with the length $L_{loop}=1.5\times10^{10}$~cm is only a fraction of a second. Thus, the observed evolution, which is two orders of magnitudes slower, requires a diffusive transport of nonthermal electrons. Given the definition of the diffusion time, $\tau_e=3L^2/(v\lambda)$, where $L\sim0.3L_{loop}$ is the size of interest, we can easily estimate the required mfp as $\lambda\sim1.5\times10^8$~cm at the energies $300-600$~keV. Now, we consider what level of magnetic turbulence is needed to provide the corresponding diffusion time of $\sim100$~sec. {We adopt} ``generic'' magnetic irregularities with a power-law spectrum

\begin{equation}
\label{Eq_B_turb_spec}
 \left<\delta B^{2}\right>_k = (\nu-1) \left<\Delta B^{2}\right> \frac{k_c^{\nu-1} dk}{k^\nu},
\end{equation}
where $\left<\Delta B^{2}\right>$ is the square of the rms magnetic field value at all $k>k_c=2\pi/L_c$. In this case, the particle mfp is estimated as \citep[][Eq. 7.106]{Fl_Topt_2013_CED}
\begin{equation}
\label{Eq_B_turb_mfp}
 \lambda \approx   \left( \frac{R_0}{L_c}\right)^{2-\nu} \frac{L_c}{\zeta_B},
\end{equation}
where $\zeta_B= {\left<\Delta B^{2}\right>}/{B^2}$.
Figure~\ref{f_diff_times}a displays the diffusion time as a function of the electron kinetic energy for different levels of the random magnetic field indicated in the panel, assuming that $L_c=10^7$~cm and $\nu=5/3$ (Kolmogorov spectrum). The required diffusion time of $\sim100$~sec is achieved for $\zeta_B\sim10^{-3}$.

Figure~\ref{f_diff_times}b displays the results of a similar exercise, but this time for the case of whistler turbulence, again bounded to the frequency range between $\omega_{Bp}$ and $\omega_{Be}$. Given the thermal plasma density in the large loop, the main scale  $L_c=v_A/\omega_{Bp}\equiv c/\omega_{pp}$, where $\omega_{pp}$ is the proton plasma frequency, is $\approx500$~cm. Here we adopted $\zeta=10^{-4}$ and considered dependencies for various spectral indices $q$ of the whistler turbulence. Thus, $\zeta=10^{-4}$ requires $q\approx2$. Given that the curves scale linearly with $\zeta$, for $q=5/3$ as in the panel (a), we need only $\zeta\approx3\times10^{-5}$ to ensure the diffusion time of the order of two minutes, as observed.

Then, in the decay phase, the apparent evolution pattern changes: there is no big change in the spatial distribution of the nonthermal electrons, while the total number of the nonthermal electrons slowly declines, which is most likely  a result of the same regime of strong diffusion but after the electron injection is off.


\section{Concluding Remarks}

The study {performed here} reveals important evolution of the physical parameters in two highly different flares--a dense single-loop flare and a multi-loop flare that involved both dense and tenuous loops. It is interesting that in the rise phase of the dense flare, the main evolution comes from the changing source geometry; primarily, increase of the flaring loop volume. In contrast, around the peak phases of both flares, the most significant evolution occurs in the nonthermal electron population, while the flux tubes do not show any noticeable change.

It is further instructive, that the observed parameter evolution requires wave-particle interaction of the electrons with turbulence, perhaps, composed of the whistler waves. In the dense flare, a modest level of turbulence is needed to support nonthermal electron population against rapid Coulomb losses in the dense loop. The same turbulence might be responsible for the decimetric component of the radio burst, likely produced by {the} RTR emission mechanism. Remarkably, the model of stochastic electron acceleration by whistler turbulence is quantitatively consistent with all available observational constraints.

In the multi-loop flare, the considered evolution of the \mw\ emission constrained mainly the nonthermal electron behavior in only one (large loop; loop \#4) of {the} four flaring loops. Therefore, we obtained a detailed 3D information about the nonthermal electron trapping in this large loop. We found that a strong diffusion regime with the nonthermal electron mfp $\lambda\sim1.5\times10^8$~cm is needed to interpret a relatively slow spatial evolution of the nonthermal electron cloud in this loop.

Therefore, our study confirms a central role of the turbulence in the electron acceleration and transport in solar flares.


\acknowledgments

This work was supported in part by NSF grants AGS-1262772  and AST-1615807, and NASA grants NNX14AC87G, NNX16AL67G, 80NSSC18K0015, and 80NSSC18K0667 to New Jersey Institute of Technology.
We would like to acknowledge the support to the 2017 NJIT-CSTR Summer High School Research Internship Program provided by the Academic Division of the Harris Corporation, which provided free IDL licenses to all participants for the duration of the program.

\bibliographystyle{apj}
\bibliography{WP_bib,solar_radio,Xray_ref,2015-06-22_flare_refs,fleishman}

\end{document}